\documentclass[sigconf]{acmart}
\usepackage{subcaption}
\usepackage{multirow}

\AtBeginDocument{%
  \providecommand\BibTeX{{%
    \normalfont B\kern-0.5em{\scshape i\kern-0.25em b}\kern-0.8em\TeX}}}

\acmPrice{15.00}
\acmISBN{978-1-4503-XXXX-X/18/06}

\begin{document}

\title{Sequential Recommendation on Temporal Proximities with Contrastive Learning and Self-Attention}

\author{Hansol Jung}
\email{sol0917@unist.ac.kr}
\affiliation{%
  \institution{Ulsan National Institute of Science and Technology}
  \city{Ulsan}
  \country{Republic of Korea}
}

\author{Hyunwoo Seo}
\authornote{Co-corresponding author}
\email{ta57xr@unist.ac.kr}
\affiliation{%
  \institution{Ulsan National Institute of Science and Technology}
  \city{Ulsan}
  \country{Republic of Korea}
}

\author{Chiehyeon Lim}
\authornote{Co-corresponding author}
\email{chlim@unist.ac.kr}
\affiliation{%
  \institution{Ulsan National Institute of Science and Technology}
  \city{Ulsan}
  \country{Republic of Korea}
}

\renewcommand{\shortauthors}{Jung, et al.}

\begin{abstract}

Sequential recommender systems identify user preferences from their past interactions to predict subsequent items optimally. Although traditional deep-learning-based models and modern trans-former-based models in previous studies capture unidirectional and bidirectional patterns within user–item interactions, the importance of temporal contexts, such as individual behavioral and societal trend patterns, remains underexplored. Notably, recent models often neglect similarities in users’ actions that occur implicitly among users during analogous timeframes—a concept we term \textit{vertical temporal proximity}. These models primarily adapt the self-attention mechanisms of the transformer to consider the temporal context in individual user actions. Meanwhile, this adaptation still remains limited in considering the \textit{horizontal temporal proximity} within item interactions, like distinguishing between subsequent item purchases within a week versus a month. To address these gaps, we propose a sequential recommendation model called TemProxRec, which includes contrastive learning and self-attention methods to consider temporal proximities both across and within user-item interactions. The proposed contrastive learning method learns representations of items selected in close temporal periods across different users to be close. Simultaneously, the proposed self-attention mechanism encodes temporal and positional contexts in a user sequence using both absolute and relative embeddings. This way, our TemProxRec accurately predicts the relevant items based on the user-item interactions within a specific timeframe. We validate this work through comprehensive experiments on TemProxRec, consistently outperforming existing models on benchmark datasets as well as showing the significance of considering the vertical and horizontal temporal proximities into sequential recommendation.

\end{abstract}




\received{20 February 2007}
\received[revised]{12 March 2009}
\received[accepted]{5 June 2009}

\maketitle
\begin{figure}[t]
    \centerline{\includegraphics[width=\columnwidth]{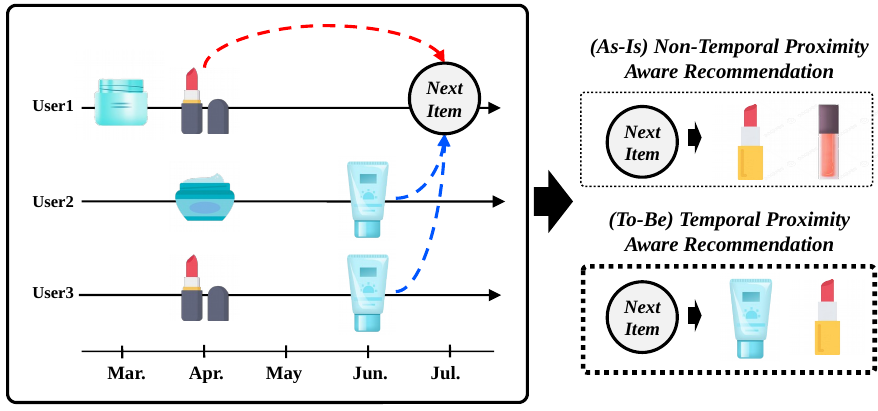}}
    \caption{(As-Is) Recommending lipstick by focusing solely on items within the user's history, (To-Be) Recommending sunscreen and lipstick by considering both items across and within users' history.
        }
    \label{Figure 1} 
\end{figure}
\section{Introduction}

Sequential recommender systems have been used in many online platforms, including online stores and online media providers, to identify item preferences of users and to lead their specific actions. These systems learn past item interactions of users to predict their subsequent item accurately. Researchers both in academia and industry have devoted significant efforts to advancing sequential recommender systems. Markov chains and recurrent neural networks traditionally have been used to capture short-term and long-term patterns in user-item interactions for recommendation \cite{rendle2010factorization, he2016fusing, lecun2015deep, wu2017recurrent}. Recently, transformer \cite{vaswani2017attention}-based models have achieved state-of-the-art performances in sequential recommendation tasks with their outstanding capabilities to represent the patterns in user-item interactions \cite{sun2019BERT4Rec, shaw2018self, li2020time, cho2020meantime, tran2023attention, rashed2022context}.

Given that the interactions occur chronologically, temporal context should be considered in sequential recommendation; consider that user preferences dynamically evolve over time and heavily rely on the temporal context either at the individual or social trends. However, while the performance of sequential recommendation has been improved through previous studies, most of them have underutilized the temporal context explicitly, implying the potential of utilizing this context to further advance sequential recommender systems. Although some recent studies start to consider the temporal context, they merely adapt the self-attention mechanisms to consider the temporal context of an individual user’s actions: TiSASRec \cite{li2020time} converts the timestamp of each user’s action into a single embedding, while MEANTIME and MOJITO \cite{cho2020meantime, tran2023attention} develop advanced attention mechanisms that incorporate multiple temporal contexts of each user’s action. 

However, such adaptation falls short in identifying similarities in user actions that occur implicitly across users during analogous timeframes, which we call \textit{vertical temporal proximity}. In addition, focusing solely on the absolute time and position of each action fails to account for the \textit{horizontal temporal proximity} within user-item interactions, like distinguishing between subsequent item purchases of a user within a week versus a month. Thus, our work delves into the concepts of vertical and horizontal temporal proximities to advance sequential recommender systems (see the recommendation example illustrated in Figure \ref{Figure 1}). Specifically, through an experiment on multiple real-world datasets, we show that the vertical and horizontal temporal proximities are critical factors in user-item interactions (see Section 2.1). This experiment demonstrates that the interactions of a user can be influenced by her/his previous actions as well as the actions of concurrent users within close timeframes.

Based on this finding, we propose to explicitly model the vertical and horizontal temporal proximities in the user-item interactions. Specifically, we develop the Temporal-Proximity-aware Recommendation model (TemProxRec), which incorporates Temporal-proximity-aware Contrastive Learning (TCL) and Multi-Head Ab-solute-Relative (MHAR) attention. The TCL method learns item representations to consider the vertical temporal proximity between the focal user’s and other users' item interactions. Note that contrastive learning is an approach to learn representations of semantically similar instances to be closer and different instances distant \cite{gutmann2012noise, mnih2013learning}. The proposed TCL method defines the items interacted with users in a predefined time window as positives, and makes the representations of the positive item pairs have similar representations. This way, the proposed TemProxRec can capture temporal dependencies among items across users’ interactions during analogous timeframes. The MHAR attention encodes the temporal and positional contexts of a user’s actions into absolute and relative embeddings and integrates them respectively at each head with item embedding. These embeddings represent pairwise relationships of actions within the user’s history based on relative time intervals and orders between items. This way, the proposed TemProxRec can recognize the sequential structure in user-item interactions while distinguishing the interactions in different timeframes. We demonstrate the validity of our work through comprehensive experiments on benchmark datasets from multiple domains.

This work is original research that presents the temporal-prox-imity-aware sequential recommendation (see Figure \ref{Figure 1}). Its academic contribution is to extend the modern sequential recommender systems literature to consider the vertical and horizontal temporal proximities into sequential recommender systems (see Section 2 for the literature review and the exploratory experiment on the temporal proximity concept). We successfully defined the problems to consider the vertical and horizontal temporal proximities as problems of time-aware contrastive learning and multi-head self-attention (see Section 3 for the proposed TemProxRec). The methodological contribution of our work was validated through comprehensive experiments (see Section 4). The results from a comparative experiment with baselines show that TemProxRec consistently outperforms recent models for sequential recommendation. The results from ablation studies further confirm the necessity of modeling and integrating vertical and horizontal proximities into sequential recommendation. In conclusion, we argue that temporal proximity is a critical yet underexplored factor that requires further investigations in the sequential recommender systems literature (see Section 5 for further discussion). For the reproduction and application of our work, we release our code on GitHub (see Appendix B.3). %https://github.com/TemProxRec. 
\begin{figure}[t]
    \centering
    \begin{subfigure}{\columnwidth}
        \centering
        \includegraphics[width=0.49\columnwidth]{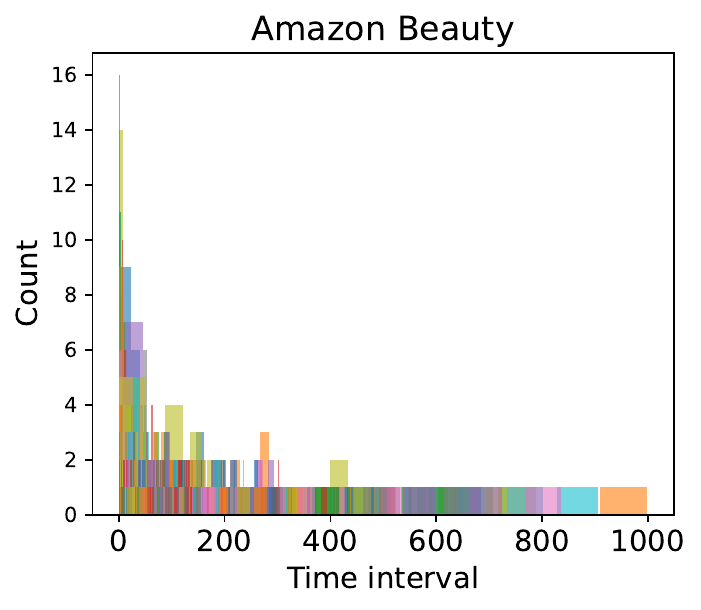}
        \includegraphics[width=0.49\columnwidth]{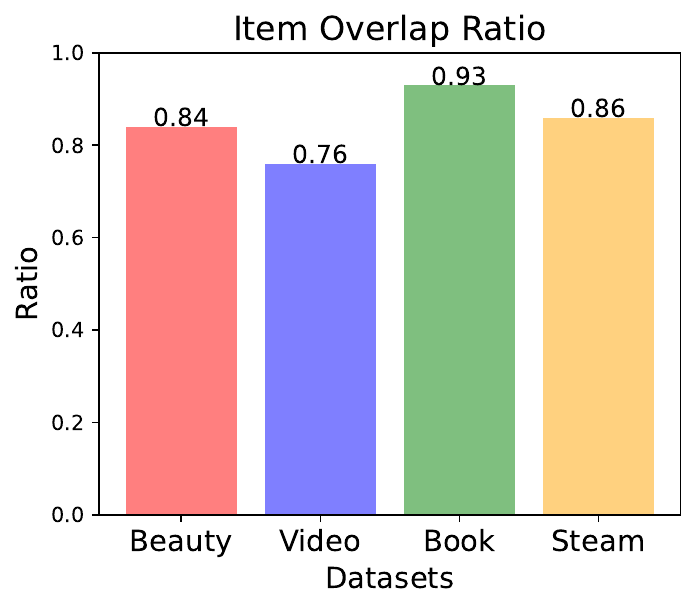}
    \end{subfigure}
    \caption{(1) Time interval distribution of sequential item pairs in Amazon Beauty dataset (except for the interval of zero), where each color indicates an item pair (2) Average item overlap ratios of four benchmark datasets}
    \label{Figure 2} 
\end{figure}
\section{Background}
\subsection{Temporal Proximity}

To demonstrate the importance of temporal proximity in sequential recommendation, we designed two experiments. In the first experiment, we calculated the time interval, measured in days, of item pairs that sequentially occur in all user-item interactions in the Amazon Beauty dataset. We found that the sequential item pairs are selected with various time intervals in between (see Figure \ref{Figure 2}-(1)). This result indicates that the position or order information of items solely does not identify the time span between items. In the second experiment, we defined the item overlap ratio $r_u$ of a user $u$ as the ratio of items that have been selected at least once by the other users within a predefined time window around the item over the total items of the user sequence. For the Amazon Beauty, Book, Video, and Steam datasets, which are the popular benchmark datasets for sequential recommendation, we calculated the average item overlap ratio for top 100 users with the most interactions given a time window with a radius of 30 days (see Figure \ref{Figure 2}-(2)). We found that the average item overlap ratio within 30 days is over 0.75 for all datasets. This result shows that there exists a tendency where items are concurrently interacted with multiple users.

The results of these experiments show that sequential item interactions within a user sequence have various levels of proximity in time (experiment 1) and that the items are selected concurrently across users in analogous timeframes (experiment 2). Thus, we hypothesize that explicitly considering the temporal proximity of items across and within users' interactions can improve the performance of sequential recommender systems. The problem formulation and the proposed model for considering the temporal proximity will be explained in detail in Section 3.

\begin{figure*}[h]
    \centerline{\includegraphics[width=\textwidth]{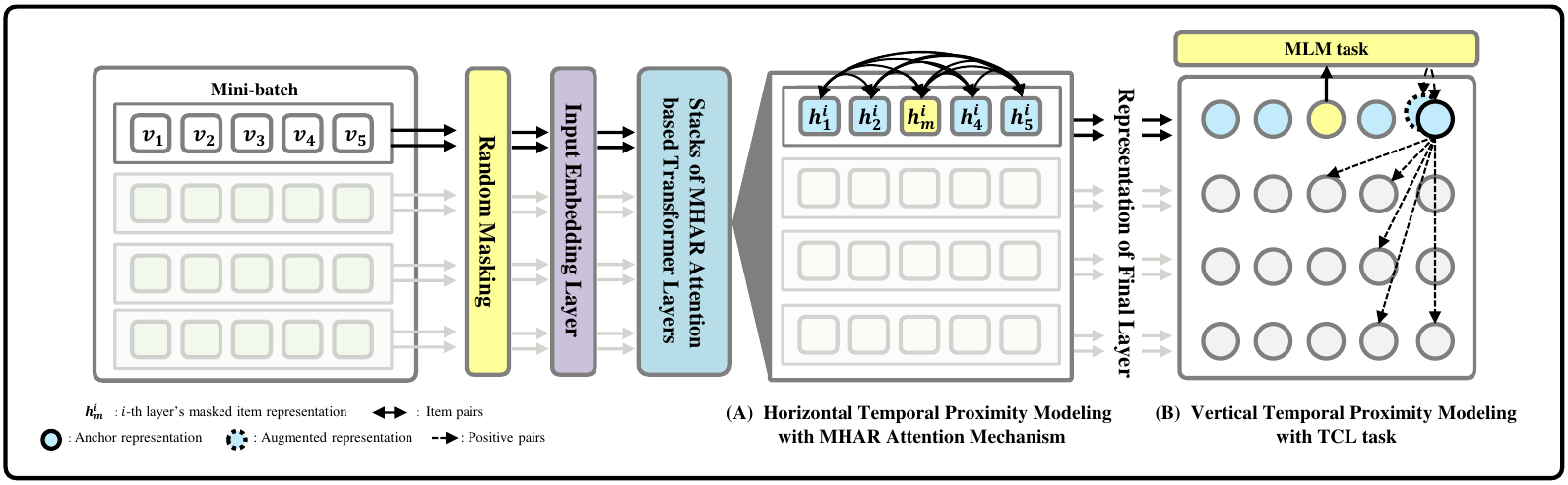}}
    \caption{Overview of TemProxRec. TemProxRec produces item representations from MHAR attention-based transformer layers. Using the final representations, TemProxRec conducts the TCL and the MLM. During the training, TemProxRec (A) performs attention within each sequence for horizontal temporal proximity, (B) contrasts items across sequences for vertical temporal proximity.}
    \label{Figure 3} 
\end{figure*}

\subsection{Sequential Recommendation}
Sequential recommendation aims to capture sequential patterns and user preferences based on the user's historical interactions. Various models have been proposed to learn the intricate sequential patterns, from traditional Markov Chain-based methods \cite{ rendle2010factorization, he2016fusing} to modern deep learning-based methods \cite{lecun2015deep, wu2017recurrent, shin2022recommendation}. Especially, recent transformer-based methods have demonstrated remarkable performance in capturing pairwise dependencies between items \cite{vaswani2017attention}. SASRec \cite{kang2018self} successfully introduced a self-attention mechanism in sequential recommendations. BERT4Rec \cite{sun2019BERT4Rec} proposed bi-directional self-attention along with a cloze task called Masked Language Modeling (MLM), which predicts randomly masked items in sequences. However, these models focus on the sequential order of items only and neglect the temporal information in sequences.

To address this limitation, TiSASRec \cite{li2020time} successfully incorporated time interval embedding into the self-attention mechanism. MEANTIME \cite{cho2020meantime} adopted multiple types of temporal embeddings within the self-attention mechanism to capture diverse temporal patterns in user-item interactions. CARCA \cite{rashed2022context} further incorporated temporal context with non-temporal context. Recently, MOJITO \cite{tran2023attention} generated multiple types of temporal embeddings and injected the concatenated temporal embedding into the mixture-based self-attention mechanism. However, while these models have paid attention to capturing temporal dependencies within a user sequence, they still overlook the temporal dependencies of items across other user sequences. Also, merging temporal embeddings solely into self-attention is insufficient for addressing the relative time differences among items, such as subsequent item interactions within a day and a month. To address these limitations, Our work incorporates time and position information, both absolute and relative, into the self-attention mechanism.

\subsection{Contrastive Learning}

Contrastive learning aims to minimize the distance between similar sample pairs and maximize the distance between dissimilar sample pairs in the latent space \cite{gutmann2012noise, mnih2013learning}. In this context, an "anchor" is a reference data point used to compare similarity, while a similar sample with the anchor is termed a "positive sample" and is paired together with the anchor to create a positive pair. Conversely, dissimilar samples are called "negative samples". The generally used loss function for contrastive learning is formulated as in Eq. \ref{Equation 1} \cite{chen2020simple}. 
\begin{equation}\label{Equation 1}
\mathcal{L}_{CTL}=\sum_{i\in N} -log\frac{\text{exp}(\text{sim}(f(x_i), f(x^+)))/\tau )}{\sum_{n=1} ^K \text{exp}(\text{sim}(f(x_i), f(x_n))/\tau )},
\end{equation}
where $f$ is encoder, $x_i$ and $x_+$ are a positive pair while $x_n$ are negative samples of $x_i$. $\tau$ is the temperature parameter.

Contrastive learning has been utilized in sequential recommendation to improve sequence representations through data-level augmentation for user sequences and employ contrastive learning on sequences. However, these approaches are unable to consider the relationship among items across user-item interactions, with respect to temporal information. To model the temporal relationship among items, we propose a novel time-aware contrastive learning on item representations.

\begin{figure*}[!t]
    \centerline{\includegraphics[width=0.98\textwidth]{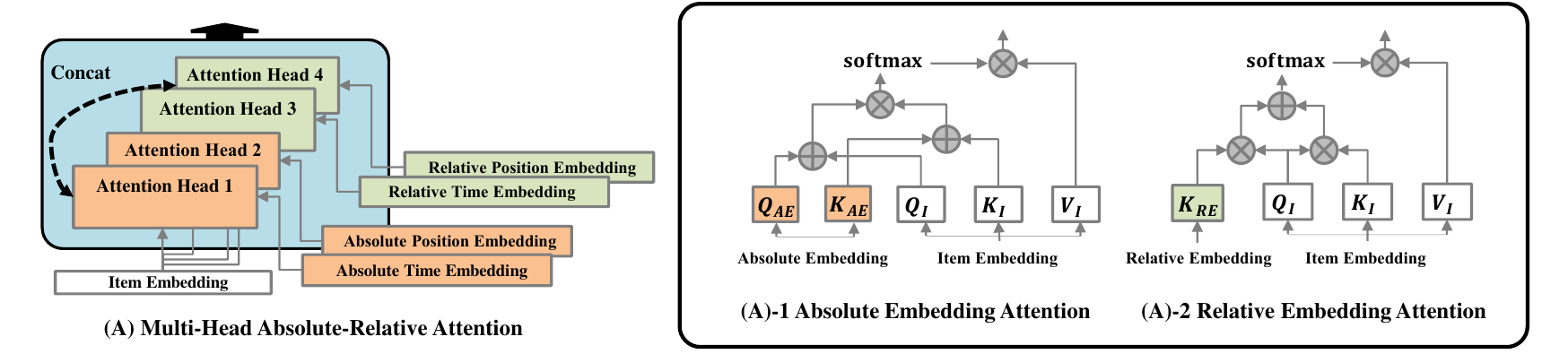}}
    \caption{Structure of the proposed Multi-Head Absolute-Relative Attention (A). Self-attention mechanism with absolute embeddings ((A)-1). Self-attention mechanism with relative embeddings ((A)-2).}
    \label{Figure 4} 
\end{figure*}

\section{Methodology}
To consider the horizontal and vertical temporal proximities in the user-item interactions, we develop a novel sequential recommendation model, TemProxRec. The overall framework is illustrated in Figure \ref{Figure 3}. TemProxRec is composed of two key components, the Multi-Head Absolute-Relative (MHAR) attention and temporal-proximity-aware contrastive learning (TCL). The MHAR attention is designed to capture the temporal proximity between items within a sequence from a horizontal axis (see Figure \ref{Figure 3}-(A)). Then, the TCL builds upon the representations from MHAR attention-based transformer layers to consider the temporal proximity among items across sequences of users in a vertical manner (see Figure \ref{Figure 3}-(B)). To learn parameters, the training process of TemProxRec includes the proposed TCL task and the MLM task, which are jointly optimized during training. 

\subsection{Problem Formulation}
Let $U$ be a set of users, $I$ a set of items, and $T$ a set of absolute timestamp values at daily intervals spanning from the initial and final times in the dataset. For each user $u\in U$, chronological item sequence is $V^u=[v_1^u,v_2^u,...,v_{|V^u|}^u $ $\mid v_i^u\in I]$, where $|V^u|$ is the number of the user $u$'s sequence. The corresponding time sequence is $T^u=[t_1^u,t_2^u,...,t_{|V^u|}^u$ $\mid t_i^u\in T]$. The sequence length is fixed at length $n$, as in previous studies \cite{kang2018self, sun2019BERT4Rec}. If the sequence is shorter than $n$, we pad a special token $[\text{PAD}]$ up to length $n$. Then, the sequences are transformed into a fixed-length sequence $v=[v_1,v_2,...,v_n]$ and $t=[t_1, t_2,...,t_n]$. The fixed position sequence is defined as in \cite{vaswani2017attention}, denoted as $p = [1, 2, ..., n]$.

The sequential recommendation problem is defined as: given sequences of user $u$, $\boldsymbol{v}=[v_{|V^u|-n+2}^u,...,v_{|V^u|}^u, [\text{MASK}]]$ and $\boldsymbol{t}=[t_{|V^u|-n+2}^u, ...,t_{|V^u|}^u, t_{pred}^u]$, the model estimates the next item $v_{pred}^u$ at timestamp $t_{pred}^u$ as output. 

\subsection{Input Embedding} 
The input embedding layer converts $v$ to hidden representations which are fed to the MHAR attention layer. We create a learnable item embedding table $M^{I} \in \mathbb{R}^{(|V|) \times d}$, where $|V|$ represents the number of unique items, and $d$ represents the hidden dimension. We then convert $v$ into the input embedding matrix $E_{v}^{I} = [M^I_{v_1}, M^I_{v_2}, \ldots, M^I_{v_n}]^\top \in \mathbb{R}^{n \times d}$. We only transform the item sequence at the input layer and utilize it as the input representation $H^{(0)}=E_{v}^{I}$. The time and position sequences are transformed and integrated with the input embedding in the MHAR attention.

\subsection{Multi-Head Absolute-Relative Attention} % Section 4.2
The MHAR attention incorporates time and position sequences with input embeddings by converting them to absolute and relative embeddings, respectively. The absolute values of sequences are transformed in absolute embeddings, while the relative differences within sequences are transformed in relative embeddings. By encoding time and position information in these ways, TemProxRec can systematically learn the horizontal temporal proximity within a sequence in detail, becoming capable of distinguishing sequential interactions within different time intervals.

The overall structure of the MHAR attention is illustrated in Figure \ref{Figure 4}-(A). As aforementioned, the MHAR attention encompasses four distinct types of embeddings at the respective heads: absolute time, absolute position, relative time, and relative position embeddings. Each head's dimension is set as $d/4$ where 4 indicates the number of heads. The self-attention mechanism combining absolute embeddings with item embeddings is depicted in Figure \ref{Figure 4}-(A)-1, while Figure \ref{Figure 4}-(A)-2 illustrates the self-attention mechanism with relative embeddings.

\subsubsection{Absolute Embeddings}
To transform the absolute time values, We create a learnable time embedding table $M^T \in \mathbb{R}^{(|T|) \times d}$, where $|T|$ represents the number of unique time values in $T$. From this look-up table, We obtains the absolute time embedding denoted as $E^{T} \in \mathbb{R}^{n \times d}$. Similarly, TemProxRec generates a learnable position embedding table $M^P \in \mathbb{R}^{n \times d}$ and obtains the absolute position embedding $E^{P} \in \mathbb{R}^{n \times d}$.

The absolute time and position embeddings are fed into the separate heads: $E^{T}$ and $E^{P}$ are separately used as inputs for $K_{AE}$ and $Q_{AE}$, which represent the key and query for absolute contexts in each head (see Figure \ref{Figure 4}-(A)-1).

\begin{figure*}[t]
    \centerline{\includegraphics[width=\textwidth]{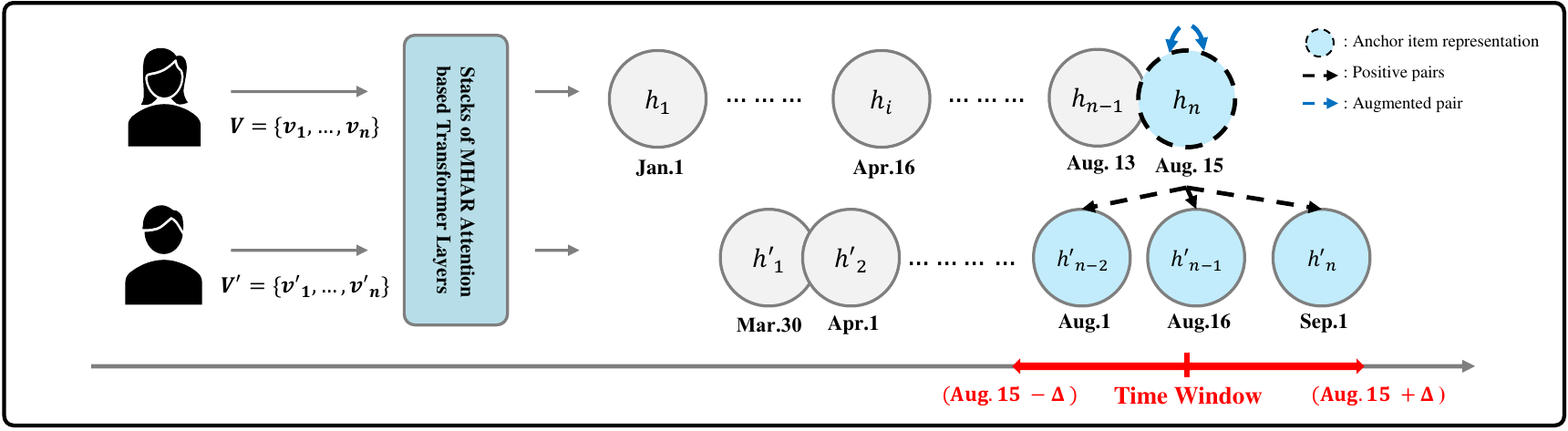}}
    \caption{Illustration of the sampling strategy for temporal-proximity-aware contrastive learning. The positive samples for an anchor are determined based on vertical temporal proximity which is accessed by a predefined time window.}
    \label{Figure 5} 
\end{figure*}

\subsubsection{Relative Embeddings}% Section 4.2.2
Relative time and position embeddings of a sequence encode relative intervals between items on time and position, respectively. Our formulation on relative embeddings is built upon a self-attention mechanism with relative positions \cite{shaw2018self}. 

To construct the relative time embedding with time sequence $\boldsymbol{t}$, we create a learnable time interval embedding table $M^{RT}\in\mathbb{R}^{k_t\times d}$, where $k_t$ represents the clipping value for the maximum time interval. We calculate the pairwise time intervals as a matrix, denoted as $TI \in\mathbb{R}^{n\times n}$. Each $TI_{ij} $ represents the time interval between the $i^{th}$ and $j^{th}$ items and is computed as Eq. \ref{Equation 2}: 
\begin{equation}\label{Equation 2}
\begin{aligned}
TI_{ij} &= \text{clip}_t(t_j - t_i, k_t), \\
\text{clip}_t(x, k) &= \min(|x|, k)
\end{aligned}
\end{equation}
Clipping a time interval by $k_t$ considers the time interval larger than $k_t$ as equally distant. We encode the absolute time differences between items through the absolute transformation in clipping operation since the order between items is considered in relative position embeddings. Finally, the relative time embedding table $M^{RT}$ converts the pairwise time interval matrix $TI$ to the relative time embedding matrix $E^{RT}\in\mathbb{R}^{n\times n\times d}$, where $E_{ij}^{RT}=M_{TI_{ij}}^{RT}$.

To encode relative position intervals into representations, we create a learnable position interval embedding table $M^{RP}\in\mathbb{R}^{(2k_{p}+1)\times d}$, where $k_{p}$ is the clipping value for the maximum position interval. Learning representations only for relative position intervals within a clipping value is known for reducing memory complexity of an embedding layer with consistent performance \cite{shaw2018self}. We calculate the pairwise relative position matrix, denoted as $PI \in\mathbb{R}^{n\times n}$. Each $PI_{ij}$ represents the positional difference between the $i^{th}$ and $j^{th}$ items, calculated as Eq. \ref{Equation 3}: 
\begin{equation}\label{Equation 3}
\begin{aligned}
PI_{ij} &= \text{clip}_{p}(j - i, k_{p}), \\
\text{clip}_{p}(x, k) &= \max(-k, \min(x, k))
\end{aligned}
\end{equation}
The pairwise relative position matrix maintains the original position differences to represent the item order. Finally, the relative position interval embedding table $M^{RP}$ converts the pairwise relative position matrix $PI$ to the relative position embedding matrix $E^{RP}\in\mathbb{R}^{n\times n\times d}$, where $E_{ij}^{RP}=M_{PI_{ij}}^{RP}$. 

As illustrated in Figure \ref{Figure 4}.(A)-2, two separate heads receive the relative time and position matrix as the input respectively, which serves as supplementary keys, denoted as $K_{RE}$, representing relative contexts. These keys are subsequently integrated with the query representation of the item. 
% Meanwhile, this self-attention mechanism, which incorporates the absolute and the relative embeddings with the item embedding is built upon the transformer - XL\cite{dai2019transformer}. For a detailed explanation, refer to \cite{dai2019transformer}. 
Meanwhile, Inspired by transformer-XL \cite{dai2019transformer}, we develop the self-attention mechanism incorporating relative embeddings, given its proven performance in adopting relative position. We further extend the mechanism to consider temporal proximity by supplementing with relative time information.

\subsubsection{Transformer Layer} % Section 4.2.3
A transformer layer of TemProxRec is composed of the MHAR attention layer and position-wise Feed-Forward Network (FFN) layer. After the four heads in the MHAR attention layer produce representations in parallel, these representations are concatenated and linearly projected. Subsequently, TemProxRec utilizes the FFN layer, which is two linear transformations with the GeLU activation in between \cite{sun2019BERT4Rec}:
\begin{equation}\label{Equation 4}
\begin{aligned}
Z^{(l+1)} &= \text{MHAR}(H^{(l)}) \\
H^{(l+1)} &= [\text{FFN}(z_1^{(l+1)})^T, ..., \text{FFN}(z_n^{(l+1)})^T] \\
\text{with } \text{MHAR}(x) &= \text{concat}(head_1, head_2, head_3, head_4)W^O \\ 
\text{FFN}(x) &= \text{GELU}(xW^1 + b^1)W^2 + b^2,
\end{aligned}
\end{equation}
where $W^O \in \mathbb{R}^{d\times d}$, $W^1 \in \mathbb{R}^{d\times 4d}$, $W^2 \in \mathbb{R}^{4d\times d}$, $b^1 \in \mathbb{R}^{4d}$, and $b^2 \in \mathbb{R}^d$ are learnable parameters. For each sublayer's output, we apply residual connection, dropout and layer normalization as in \cite{sun2019BERT4Rec}. 
Finally, we stack $L$ transformer layers to obtain the final hidden representations $H^{(L)}=[h^{(L)}_1, h^{(L)}_2, ..., h^{(L)}_n]$.

\subsection{Temporal-proximity-aware Contrastive Learning} 
While the MHAR attention captures the horizontal temporal proximity within a user's interactions with items, we propose a novel training method, temporal-proximity-aware contrastive learning (TCL), for modeling and learning the vertical temporal proximity of item interactions across users. The TCL method performs time-aware contrastive learning on item representations from $H^{(L)}$. This way, the TCL allows the representations of items that are shared among users in close timeframes to become similar.

Specifically, the TCL method samples contrastive pairs (i.e., positive and negative samples) based on the temporal proximity between an anchor and other items (see Figure \ref{Figure 5}). An anchor is designated as the last item of each sequence from the minibatch. To assess the temporal proximity between an anchor and the items in sequences of other users, we define a time window on each anchor, centered at the anchor's timestamp $t_i$ with radius $\Delta$, $[t_i-\Delta, t_i+\Delta]$. Items that the other users selected within the time window are regarded as positive samples. If items are selected outside the time window, they are regarded as negative samples. The positive samples represent items that are likely to be co-interacted among concurrent users in adjacent timeframes. In some cases, an anchor has no positive instances if not a single user in the minibatch interacted with items within the time window. To address this issue, the TCL method generates a pseudo-positive instance using a dropout strategy like \cite{zhang2022frequency}. By inputting the sequence into the model twice with different dropout masks, we can obtain augmented representation for each anchor item, ensuring the generation of at least one positive sample. This way, positive instances of an anchor comprise the representations of items within the time window as well as the representation of the pseudo-positive instance. Based on the sampled contrastive pairs, the following proposed loss function is utilized for the optimization of the TCL:
\begin{equation}\label{Equation 5}
\mathcal{L}_{TCL}= \sum_{i\in U} \frac{1}{|pos_i|}\sum_{p\in pos_i} -\text{log}\frac{\text{exp}(\text{sim}(h_i, h_p))/\tau )}{\sum_{n=1}^{K_i} \text{exp}(\text{sim}(h_i, h_n)/\tau )},
\end{equation}
where $sim(·,·)$ is cosine similarity and $h_i$ and $h_p$ are the representation of the anchor and positives; $h_n$ is the representation of the negatives. $|pos_i|$ is the number of the positives of $h_i$ and $\tau$ is the temperature. By optimizing $\mathcal{L}_{TCL}$, representations of items selected from concurrent users at similar timeframes become closely aligned, reflecting the vertical temporal proximity. The TCL method is related to supervised contrastive learning which samples contrastive pairs through external information \cite{khosla2020supervised}. It employs the temporal information (e.g., timestamp) of each item for defining contrastive pairs based on their temporal proximity. Meanwhile, the TCL method is the contrastive learning on item representations to learn the vertical temporal proximity among them, different from previous studies that apply contrastive learning on sequence representations (refer to Section 2.3).

\subsection{Optimizing TemProxRec} % Section 4.4

In the training stage, we perform two tasks: the conventional MLM task \cite{devlin2018bert, sun2019BERT4Rec} and our proposed TCL task. For the former task, we randomly replace a proportion $\rho$ of items in the input sequence with $[\text{MASK}]$ and predict these masked items using feed-forward networks \cite{sun2019BERT4Rec}. The MLM loss function is the negative log-likelihood:
\begin{equation}\label{Equation 6}
\mathcal{L}_{MLM}=\sum_{u\in U}\sum_{v_{m}\,is\,masked}-\text{log} P(v_{m} = v^*_{m}| \hat{\boldsymbol{v}}^u)
\end{equation}
where $\hat{\boldsymbol{v}}^u$ is the masked item sequence $\boldsymbol{v}^u$ of user $u$, $v_{m}$ is a predicted item and $v^*_{m}$ is the true item. As a result, TemProxRec is optimized through the MLM and proposed TCL tasks with the composite loss function, denoted as $\mathcal{L}$:
\begin{equation}\label{6}
\mathcal{L}=\mathcal{L}_{MLM}+\lambda \mathcal{L}_{TCL}
\end{equation}
where $\lambda$ controls the weight of the TCL task in training TemProxRec.

% Overall Experiment Result Table
\begin{table*}[ht]
\caption{Sequential recommendation performance of TemProxRec and other baselines on all datasets. The best scores and the second scores are denoted in bold and underlined. Improvements over baselines are shown in the last column.}
\label{table1}
\resizebox{\textwidth}{!}{%
\begin{tabular}{llccccccl}
\hline
Data set  &
  Metric &
  \multicolumn{1}{l}{SASRec} & 
  \multicolumn{1}{l}{BERT4Rec} & 
  \multicolumn{1}{l}{TiSASRrec} &
  \multicolumn{1}{l}{MEANTIME} &
  \multicolumn{1}{l}{MOJITO} &
  \multicolumn{1}{l}{\textbf{TemProxRec}} &
  Improv. \\ \hline
\multirow{2}{*}{Beauty} & HR@10 & 0.477 & 0.498 & 0.459 & 0.512 & \underline{0.516} & \textbf{0.535} & +3.68\%\\
                               & NDCG@10 & 0.317 & 0.335 & 0.300 & 0.344 & \underline{0.346} & \textbf{0.365} & +5.49\% \\
\multirow{2}{*}{Book} & HR@10 & 0.825 & 0.835 & 0.833 & 0.846 & \underline{0.848} & \textbf{0.860} & +1.42\% \\
                               & NDCG@10   & 0.600 & 0.627 & 0.621 & \underline{0.645} & 0.634 & \textbf{0.675} & +4.65\% \\
\multirow{2}{*}{Video Games}   & HR@10 & 0.668 & 0.668 & 0.643 & 0.685 & \underline{0.692} & \textbf{0.723} &  +4.48\%\\
                               & NDCG@10   & 0.437 & 0.450 & 0.419 & 0.471 & \underline{0.470} & \textbf{0.507} & +7.87\% \\

\multirow{2}{*}{Steam}         & HR@10 & 0.750 & 0.747 & 0.741 & 0.775 & \underline{0.779} & \textbf{0.790} & +0.14\% \\
                               & NDCG@10   & 0.515 & 0.542 & 0.502 & 0.546 & \underline{0.546} & \textbf{0.570} & +4.40\% \\ \hline

\end{tabular}}
\end{table*}

\section{ Experiment }

\subsection{Experimental Setting} 
\subsubsection{Dataset} 
We evaluated our proposed model on four real-world benchmark datasets from different domains and with varying sparsity levels, all of which include timestamp information.
\begin{itemize}
\item Amazon Beauty, Book, Video\footnote{\url{http://jmcauley.ucsd.edu/data/amazon/}}: A series of datasets on product reviews crawled from \textit{Amazon.com}. This dataset is introduced in \cite{he2016ups} and is highly sparse. We selected three popular categories, namely, "Beauty", "Books", and "Video Games". 
\item Steam\footnote{\url{https://cseweb.ucsd.edu/~jmcauley/datasets.html\#steam_data}}: A game item dataset including information such as user’s play hours, media score, and developer details, which is collected from Steam, a large online video game distribution platform. This dataset is introduced in SASRec \cite{kang2018self}.
\end{itemize}

To preprocess the dataset, we followed the preprocessing procedure commonly used in the literature \cite{rendle2010factorizing, kang2018self, tang2018personalized, sun2019BERT4Rec}. We converted each dataset into an implicit dataset by treating ratings and reviews as user-item interactions. Then, we group the interactions by unique user IDs to form a sequence and sort it based on the timestamp. To ensure the dataset quality, we typically filter out users and items that appear less than five times. In addition, for Book dataset, we applied filtering criteria described in the MOJITO \cite{tran2023attention} paper for a fair comparison, removing users and items that occur less than 30 and 20 times, respectively. For Steam dataset, we filtered out users and items that appear less than 10 and 5 times, respectively. Finally, we sampled data over a 3- to 4-year period, specifically when timestamp information was consistently available. Given the space limitations, the final dataset statistics are described in Appendix A.

\subsubsection{Evaluation}
For each user sequence, the last item in the sequence was used for the test, while the item just before the last one for the validation. To ensure a fair and simple evaluation, we adopted the negative sampling strategy in \cite{sun2019BERT4Rec}. For each user $u$, we randomly select about 100 items they haven't interacted with and rank them alongside the ground-truth item. We used two measures widely used for the evaluation of ranked item lists: \emph{Hit Ratio} (HR@K) and \emph{Normalized Discounted Cumulative Gain} (NDCG@K). We set K to 10, meaning that the model recommends 10 items for each user. After evaluating the recommendation performance of TemProxRec, we conducted ablation studies presented in Section 4.3 to assess the significance of the TCL and the MHAR attention.

\subsubsection{Baselines}
We compared our TemProxRec with state-of-the-art baselines, including both non-temporal and temporal sequential recommendation models. For non-temporal baselines, we selected SASRec \cite{kang2018self} and BERT4Rec \cite{sun2019BERT4Rec}, which are well-known transformer-based methods. To assess the impact of modeling temporal proximity, we selected the following temporal baselines, which are recent sequential recommender systems incorporating temporal information: TiSASRec \cite{li2020time}, MEANTIME \cite{cho2020meantime}, and MOJITO \cite{tran2023attention}. MOJITO is the most recent state-of-the-art model. For a fair comparison, we excluded CARCA \cite{rashed2022context} from temporal baselines as it utilizes additional non-temporal context information such as the product category. For the implementation of these baselines, please refer to the Appendix B.3.

\subsubsection{Parameter Setting} 
For the parameter setting, we fixed the batch size and maximum length of sequence across all models as 128 and 50, respectively. For common hyperparameters, all models were fairly tuned through a grid search on validation items. We considered the hidden dimension in \{16, 32, 64, 128\}, weight decay in \{0, 0.00001\}, learning rate in \{0.001, 0.0001\}, and dropout rate in \{0.1, 0.2, 0.3, 0.4 0.5\}. For other parameters unique in each model, we followed the guidelines in the original papers. For TemProxRec, all parameters were initialized using the normal distribution in range [-0.02, 0.02]. We trained our model using Adam \cite{kingma2014adam} with a learning rate of 0.001. Additionally, we tuned the temperature $\tau$ for TCL in \{0.05, 0.1\}, the radius of a time window $\Delta$ in \{7, 15, 30, 60, 100\}, weight for the TCL task $\lambda$ in \{0.1, 0.2, 0.3, 0.4, 0.5\}. We tuned the clipping value for maximum time interval $k_{t}$ in \{128,256,512,1024\}, and fixed the clipping value for maximum position interval $k_{p}$ as two as suggested in \cite{shaw2018self}.

\begin{table}[t]
\centering
\caption{Ablation study for effects of the TCL and the MHAR attention on all datasets (NDCG@1O).}
\label{table2}
\resizebox{\columnwidth}{!}{%
\begin{tabular}{ccccc}
\hline
Architecture & \multicolumn{4}{c}{Dataset}\\
\multicolumn{1}{l}{} & Beauty  & Book & Video & Steam \\
\hline
TemProxRec & \textbf{0.365} & \textbf{0.675} & \textbf{0.507} & \textbf{0.570} \\
\hline
(1) w/o TCL & 0.357 & 0.660 & 0.478 & 0.557 \\
(2) w/o abs MHAR & 0.350 & 0.651 & 0.454 & 0.546 \\
(3) w/o rel MHAR & 0.352 & 0.665 & 0.484 & 0.553 \\
(4) w/o MHAR & 0.347 & 0.630 & 0.457 & 0.537 \\
\hline%
\end{tabular}
}
\end{table}

\subsection{Performance Comparison}  
Table \ref{table1} reports all comparison results between TemProxRec and the baselines. This table shows that TemProxRec improves over all baselines on all metrics and datasets. On average, TemProxRec achieves 2.43\% and 5.6\% improvements over the strongest baselines in HR@10 and NDCG@10, respectively.

In comparing the baselines, the temporal baselines in most cases outperformed the non-temporal baselines. This confirms that consideration of temporal contexts can boost the performance of sequential recommendation in various domains. However, TiSASRec could not show impressive improvement over non-temporal baselines and even exhibited lower performance in Beauty, Video, and Steam datasets. This result shows that using a single time embedding may cause the information bottleneck problem, decreasing the recommendation performance. This finding indicates the necessity of a systematic modeling strategy for considering temporal context. From this perspective, MEANTIME and MOJITO outperform SASRec and BERT4Rec by a large margin. MEANTIME encodes temporal contexts into multiple embeddings and integrates them in different heads. MOJITO develops mixture-based attention mechanism to incorporate multiple temporal contexts and shows better performance than MEANTIME.

Compared to all these models, TemProxRec achieves the best performance. The TCL method successfully captures the vertical temporal proximity of items among concurrent users. Simultaneously, the MHAR attention models the horizontal temporal proximity within a user’s past interactions. While the baselines show good performance, they overlook the influence of these temporal proximities, primarily focusing on sequential dependencies in each user history only. Therefore, we argue that both vertical and horizontal proximities are the key drivers of the superior performance of TemProxRec.

\subsection{Ablation Study} 
\subsubsection{Effects of the proposed TCL and MHAR attention}
To measure the effects of the main components of TemProxRec, we conducted an ablation study for all datasets with the evaluation metric of NDCG@10. The variants are listed as follows: 

\begin{enumerate}
\item w/o TCL: TemProxRec removing the TCL.
\item w/o abs MHAR: TemProxRec removing absolute time and position embedding attentions.
\item w/o rel MHAR: TemProxRec removing relative time and position embedding attentions.
\item w/o MHAR: TemProxRec replacing all MHAR attentions with the canonical multi-head self-attention in \cite{vaswani2017attention}. 

\end{enumerate}

Table \ref{table2} shows that the removal of any single component leads to a decrease in performance across the datasets. Specifically, the performance drop of (1) w/o TCL suggests that the TCL method considering the vertical temporal proximity improves the sequential recommendation performance. Moreover, TemProxRec without the vertical temporal proximity still achieves higher performance than MEAMTIME and MOJITO (see Table \ref{table1}). This result demonstrates that our MHAR attention is also capable of leveraging the time and position information to represent the temporal context within a user's history. Based on these results, we argue previous studies could not fully exploit temporal contexts for sequential recommendation.

Removing the absolute or relative embedding attentions respectively decreases the performance. Especially, (2) w/o abs MHAR shows a larger drop in performance compared to (3) w/o rel MHAR. This result indicates that the absolute time and position information is more significant than the information on relative contextual differences between items. However, TemProxRec with full MHAR attention surpasses both (2) and (3) in performance. This fact suggests that the combination of (2) and (3) allows TemProxRec to effectively model the granularity of temporal contexts within a user’s item interactions. As such, removing both absolute and relative embeddings worsens the model performance; see the largest performance drop in (4) w/o MHAR. This result comes from ignoring the temporal information in sequential recommendation.

\subsubsection{Application of TCL to Other Models}

In this section, we evaluate the applicability of the proposed TCL method to advance transformer-based sequential recommendation models. We selected models to adopt the TCL task as follows: (1) The canonical transformer architecture in \cite{vaswani2017attention}, (2) Transformer architecture with temporal embedding (Transformer-T). (3) Transformer architecture in MEANTIME which utilizes multiple temporal embeddings. (4) Transformer architecture in TemProxRec. We implemented (1) and (2) by replacing the MHAR attention of TemProxRec with the basic self-attention in \cite{vaswani2017attention} and we transformed the timestamps into embedding for the temporal embedding in (2). For (3), we followed the same time encoding strategy in MEANTIME. To verify the effect of the TCL method, we added the loss function of TCL to the loss function of each case during the training. For a fair comparison, we set the same hyperparameters for the time interval and weight of the TCL task as 60 and 0.3.

\begin{figure}[t]
    \centering
    \begin{subfigure}{\columnwidth}
        \centering
        \includegraphics[width=0.49\columnwidth]{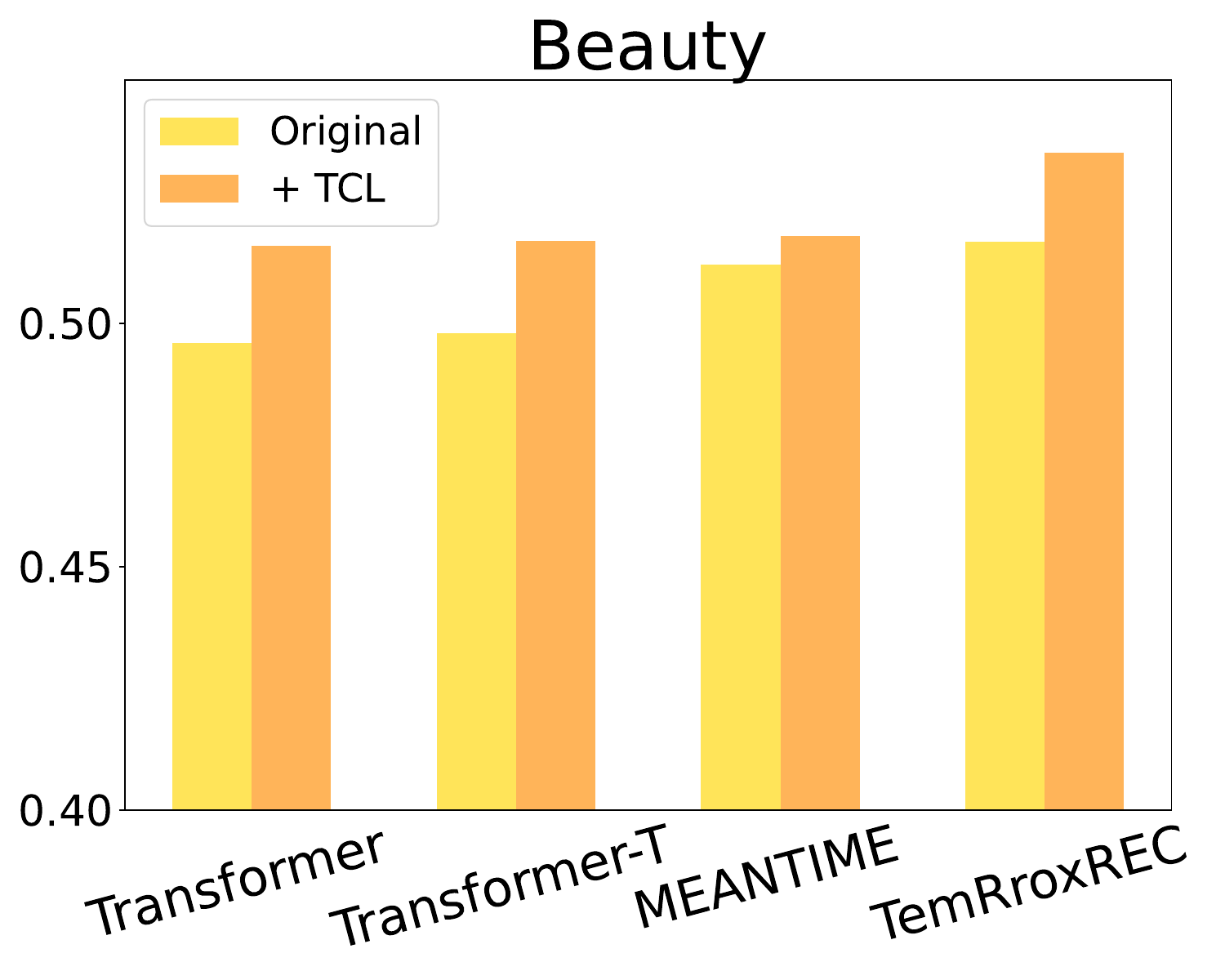}
        \includegraphics[width=0.49\columnwidth]{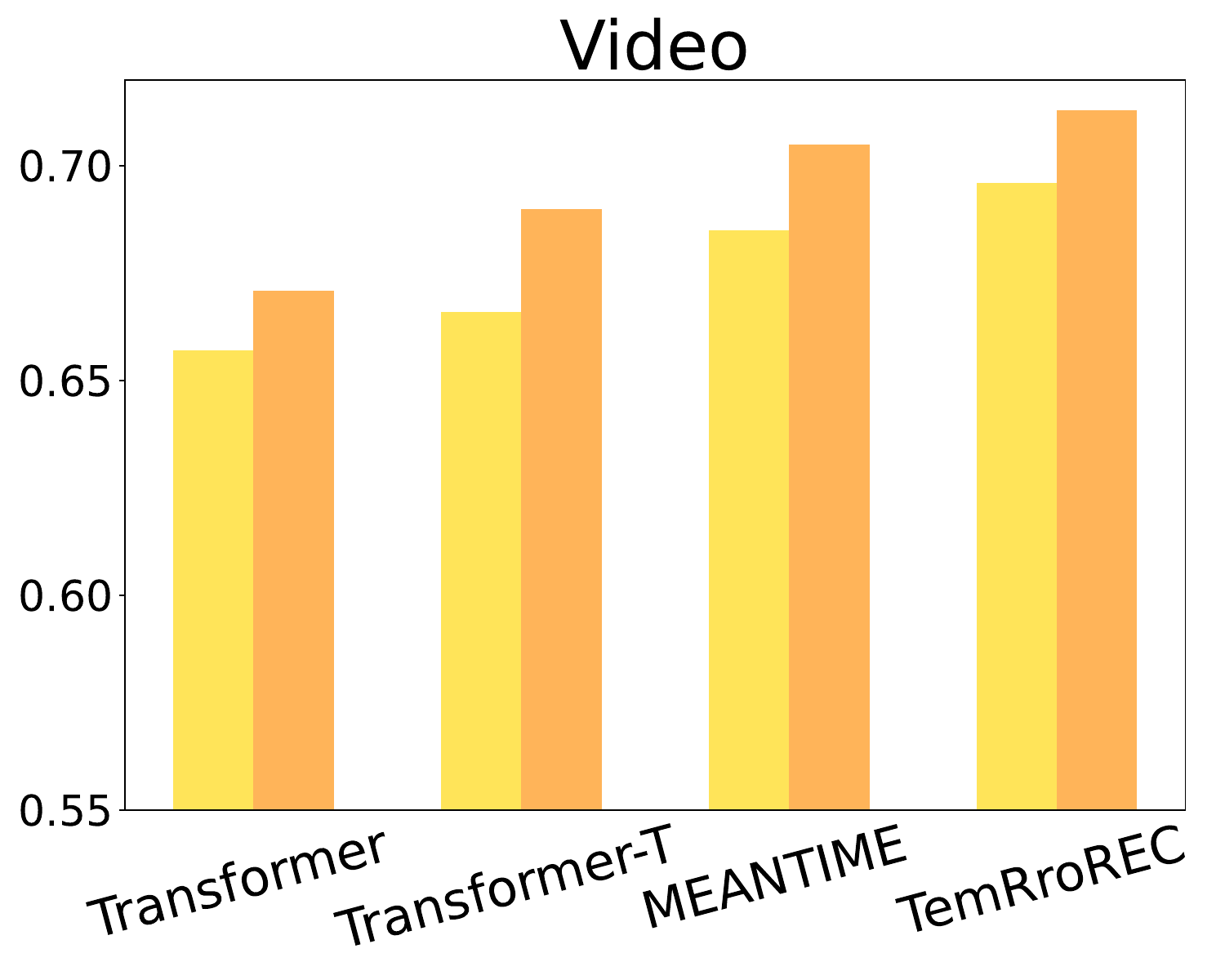}
    \end{subfigure}
    \caption{Performance comparison for impact of the TCL on Beauty and Video datasets (HR@10).}
    \label{Figure 6} 
\end{figure}

The performance of the variants on Beauty and Video datasets are shown in Figure \ref{Figure 6}. We found that the TCL task improves the performance of all models. This result, demonstrating TCL’s positive impacts on various transformer-based methods, indicates that vertical temporal proximity is an essential concept for advancing sequential recommendation. In addition, TemProxRec outperformed all other models jointly trained with the TCL task. This result indicates the TCL and the MHAR attention, combined together, successfully reflect temporal contexts with high granularity. Given all these consistent results, we argue TemProxRec is an effective, solid model to consider the horizontal and vertical temporal proximities in sequential recommendation.

\begin{figure}[t]
    \centering
    \begin{subfigure}{\columnwidth}
        \centering
        \includegraphics[width=0.49\columnwidth]{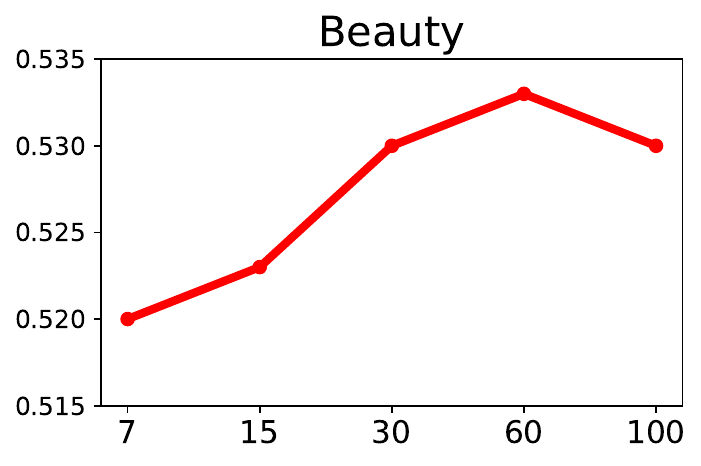}
        \includegraphics[width=0.49\columnwidth]{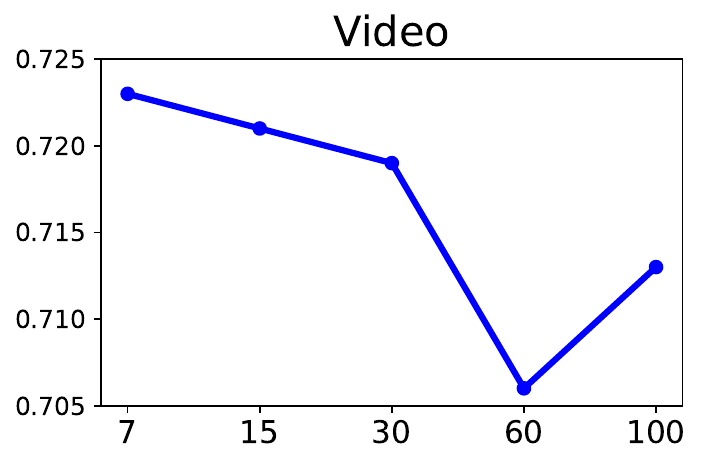}        \includegraphics[width=0.49\columnwidth]{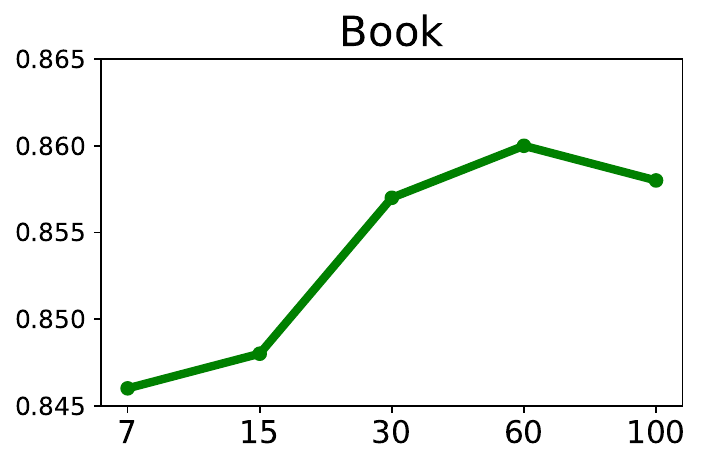}
        \includegraphics[width=0.49\columnwidth]{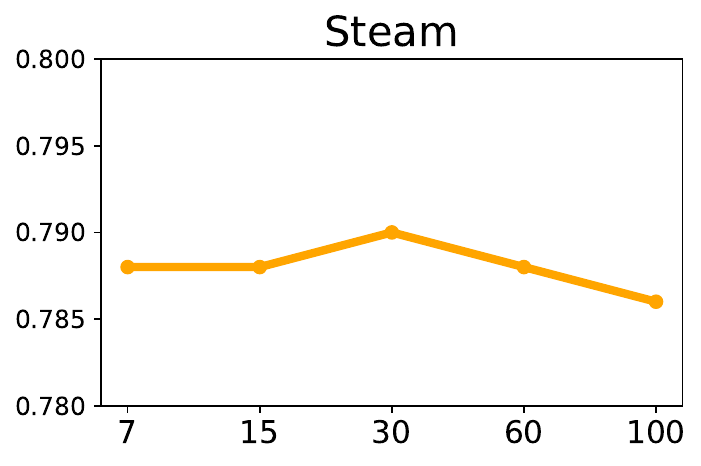}
    \end{subfigure}
    \caption{Performance for different $\Delta$ values in the TCL task on all datasets (HR@10).}
    \label{Figure 7} 
\end{figure}

\subsubsection{Parameter Sensitivity on TCL}  

When performing the TCL task, the parameter $\Delta$ defines the radius of a time window, which determines the level of temporal proximity among items. Based on $\Delta$, the TCL method allows items in close temporal proximity to have similar representation. To identify the optimal time window for each dataset and investigate the sensitivity of $\Delta$, we tested several values for $\Delta$: \{7, 15, 30, 60, 100\}, which corresponds respectively to one week, half a month, one month, two months (approximating the concept of seasonality), and more. 

Figure \ref{Figure 7} shows the values of HR@10 for different $\Delta$ values on all datasets. For Beauty and Book datasets, performance shows an improvement as the $\Delta$ value increases up to 60, but it begins to decline beyond this point. In contrast, Video achieves its best performance with shorter $\Delta$ at 7. Performance on Steam reaches its peak at 30 but is less responsive to the value with smaller variations compared to the other datasets.

These results indicate that the $\Delta$ differs depending on the item domains. As the vertical temporal proximity considers the co-occurrence among items in the close time period, the unique consumption patterns of items in each dataset influence the determination of optimal $\Delta$. For example, seasonal factors may appear to play a significant role in forming shared interests across users of Beauty and Book. On the other hand, users' interests change more rapidly, and the best performance is achieved with the shorter $\Delta$ in Video. Thus, it is important to decide the proper time window based on the unique characteristics of the user and item domain in question to achieve optimal performance of sequential recommendation with TemProxRec.

\section{Conclusion} 
In this paper, we introduced TemProxRec, a novel sequential recommender system that considers the concepts of vertical and horizontal temporal proximities in use-item interactions. Specifically, we proposed the Temporal-proximity-aware Contrastive Learning (TCL) method and Multi-Head Absolute-Relative (MHAR) attention, leading to effective modeling of the horizontal temporal proximity within a user's item interactions as well as vertical temporal proximity across item interactions of multiple users. We demonstrated the state-of-the-art performance of our TemProxRec and the significance of considering the temporal proximity concepts in sequential recommendation through a series of experiments. The ablation studies show the individual contributions of the TCL and the MHAR attention to consider the vertical and horizontal temporal proximities, respectively. 

Meanwhile, TemProxRec's intended approach is to model the temporal proximity for capturing the similarities in users’ actions. In future work, we will refine the approach to define the temporal proximities in a more comprehensive manner. For example, the advanced TemProxRec can incorporate additional contexts, such as the frequency of a user's actions and the similarity of item attributes, to give different weights to concurrent items. This way, TemProxRec will be able to infer the user's subsequent actions based on a comprehensive understanding of the temporal proximity in user-item interactions.

\bibliographystyle{ACM-Reference-Format}
\bibliography{sample-base}

\appendix
\section{Dataset statistics} 
Table \ref{table0} summarizes the statistics of the datasets after preprocessing. As indicated in the table, our datasets exhibit variations in terms of their average length and sparsity. We selected the interactions with complete timestamps; we selected records from 2011 to 2014 for Beauty and Video, from 2011 to 2013 for Book, and from 2014 to 2016 for Steam.

\begin{table}[hbt!]
\centering
\caption{Dataset statistics (after preprocessing)}
\label{table0}
\resizebox{\columnwidth}{!}{%
\begin{tabular}{lccccc}
\hline
Dataset & \#users & \#items & \#actions & Avg.length & Sparsity\\
\hline
Beauty & 20508 & 11382 & 179,580 & 6.76 & 99.92\%\\
Book & 19745 & 31671 & 932,252 & 45.21 & 99.85\%\\
Video Games & 12787 & 5846 & 107,940 &  6.44 & 99.86\%\\
Steam & 16181 & 7451 & 289,826 & 15.91 & 99.76\%\\
\hline
\end{tabular}%
}
\end{table}

\section{Parameter sensitivity}
\subsection{Clipping Value for the Maximum Time Interval}
\begin{figure}[hbt!]
    \centerline{\includegraphics[width=0.9\columnwidth]{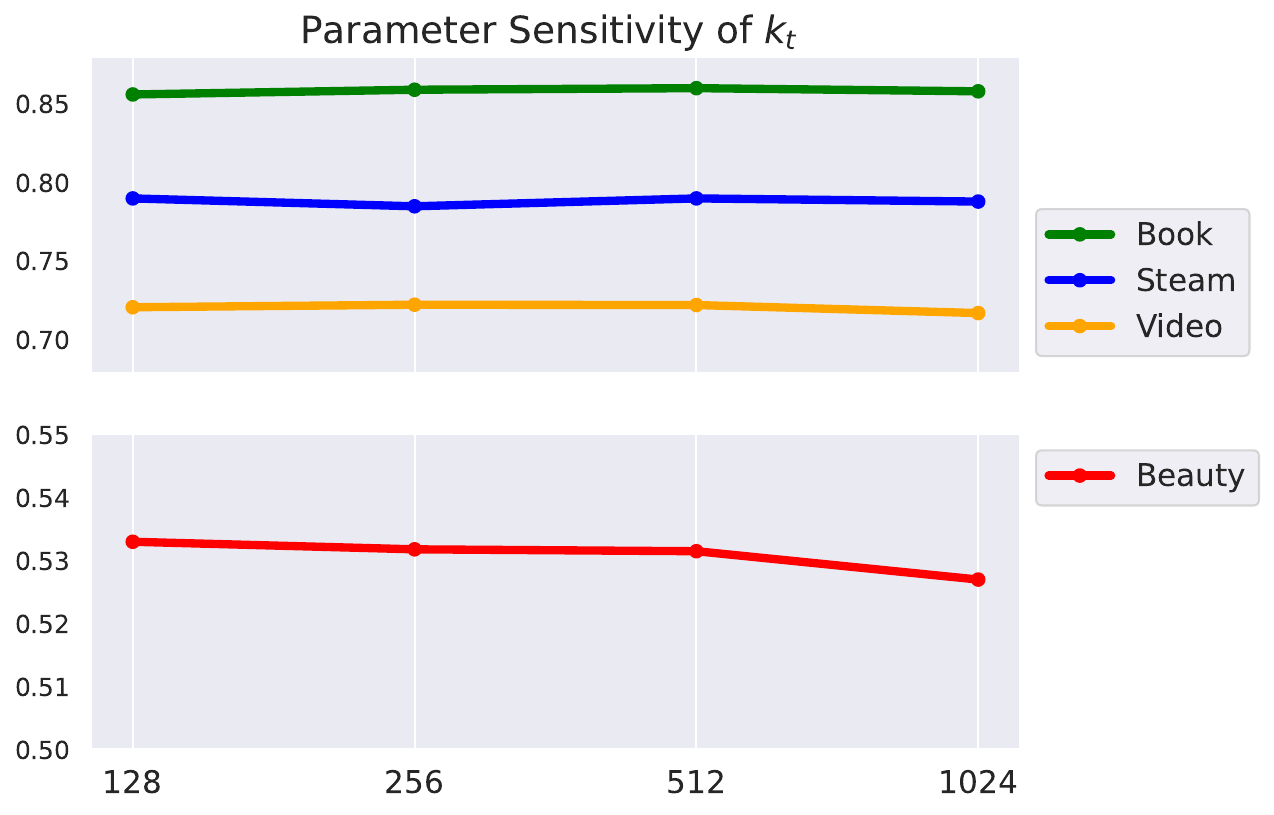}}
    \caption{Performance for different $k_t$ values (HR@10).}
    \label{Figure 8} 
\end{figure}

The clipping value for the maximum time interval $k_t$ determines the maximum value of time intervals between two items TemProxRec considers. We performed a sensitivity analysis on this hyperparameter (Figure \ref{Figure 8}). The optimal clipping values are 256 for Book and Video, 128 for Beauty, and 512 for Steam. However, the performance differences among various values are not significant for all datasets, which indicates TemProxRec is robust across the clipping values for the maximum time interval.

\subsection{Weight of the TCL Loss}
\begin{figure}[hbt!]
    \centerline{\includegraphics[width=0.9\columnwidth]{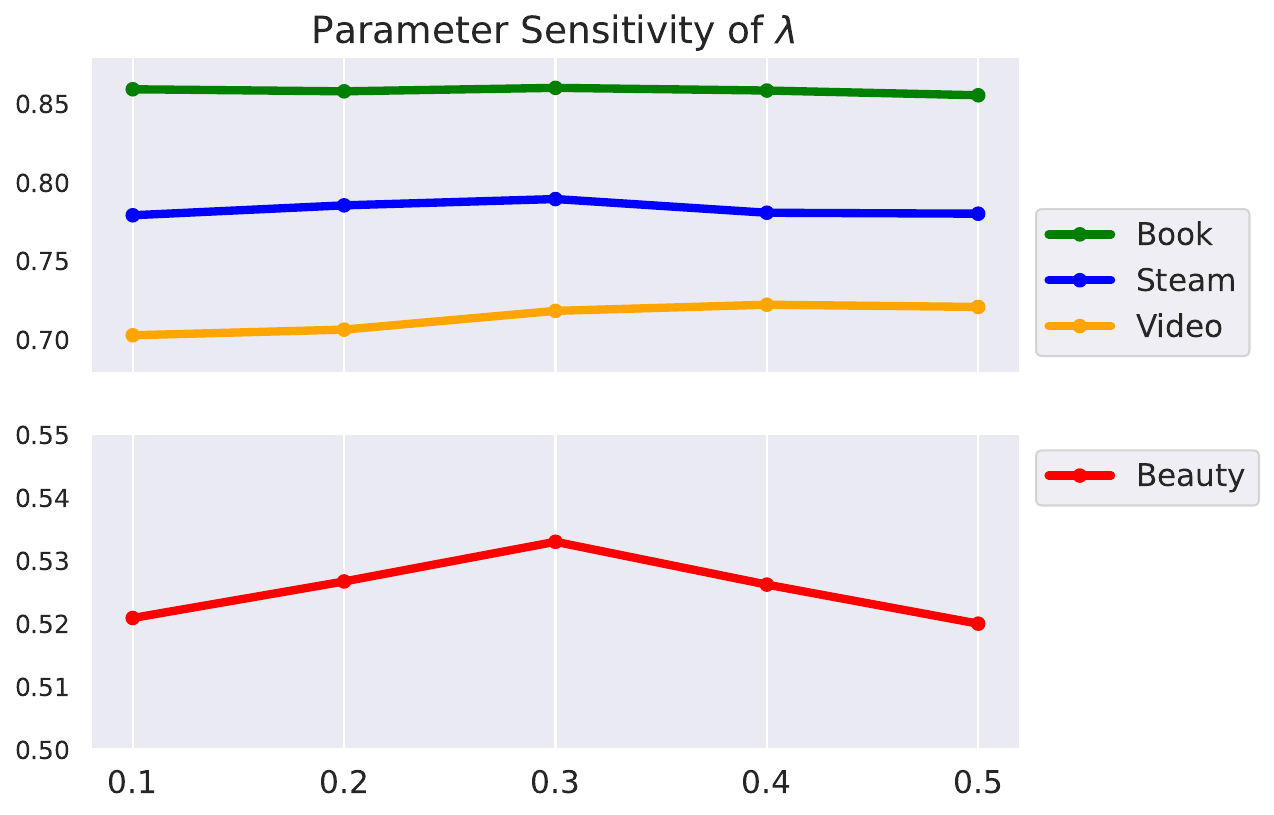}}
    \caption{Performance for different $\lambda$ values, weight for TCL loss (HR@10).}
    \label{Figure 9} 
\end{figure}

The weight for the TCL loss $\lambda$ balances the effects of the MLM and the TCL tasks in the overall loss function, $L=L_{MLM}+\lambda L_{TCL}$. We also performed a sensitivity analysis on $\lambda$ (Figure \ref{Figure 9}). In the range \{0.1, 0.2, 0.3, 0.4, 0.5\}, the optimal weight is 0.3 for Steam, Book, and Beauty datasets and 0.4 for Video dataset. 

\subsection{Experimental Reproduction}
We implemented TemProxRec with PyTorch \cite{paszke2019pytorch}. The source code and the optimal parameter values are available in our anonymous GitHub repository\footnote{\url{https://github.com/TemProxRec}}. MEANTIME\footnote{\url{https://github.com/SungMinCho/MEANTIME}}, MOJITO\footnote{\url{https://github.com/deezer/sigir23-mojito}} were implemented by the authors. We implemented BERT4Rec, SASRec, TiSASRec with PyTorch.

\end{document}